\documentclass[final,3p,times,twocolumn]{elsarticle}

\usepackage{lineno}

\usepackage{graphicx}
\usepackage{dcolumn}
\usepackage{bm}
\usepackage{color}
\usepackage{hyperref}
\usepackage{multirow}
\usepackage{amssymb}
\usepackage{amsmath}

\usepackage{amsthm}
\usepackage{bm}
\usepackage{booktabs}
\usepackage{graphicx}
\usepackage{epsfig}
\usepackage{epstopdf}
\usepackage{mathtools}
\usepackage{cuted}
\usepackage{upgreek}
\usepackage{textgreek}

\begin{document}

\begin{frontmatter}

\title{Evolution of cooperation in multi-agent systems with time-varying tags, \\ multiple strategies, and heterogeneous invasion dynamics}

\author[a]{Wonhee Jeong\corref{cor1}}
\author[b]{Tarik Hadzibeganovic\corref{cor1}}
\author[a]{Unjong Yu}
\ead{uyu@gist.ac.kr}

\address[a]{Department of Physics and Photon Science, Gwangju Institute of Science and Technology, Gwangju 61005, South Korea}
\address[b]{Department of Psychology, Faculty of Natural Sciences, University of Graz, Graz 8010, Austria}
\cortext[cor1]{These authors contributed equally to this work.}

\begin{abstract}

The degree of cooperation that can be attained in an open dynamic system fundamentally depends upon information distributed across its components. Yet in an environment with rapidly enlarging complexity, this information may need to change adaptively to enable not only cooperative interactions but also the mere survival of an organism. Combining the methods of evolutionary game theory, agent-based simulation, and statistical physics, we develop a model of the evolution of cooperation in an ageing population of artificial decision makers playing spatial tag-mediated prisoner's dilemma games with their ingroup neighbors and with genetically unrelated immigrant agents. We study the behavior of this model in the presence of four conditional and two unconditional strategies, and we introduce the concept of time-varying tags such that the phenotypic features of 'new' agents that invade the system from the outside can change into 'approved' following variable approval times. In a series of systematic Monte Carlo simulations, we observed that ingroup-biased ethnocentric cooperation can dominate only at low costs and short approval times. In the standard 4-strategy model with fixed tags, we identified a critical cost $c_{\mathrm{crit}}$ above which cooperation transitioned abruptly into the phase of pure defection, revealing remarkable fragility of ingroup-biased generosity. In our generalized 6-strategy model with time-varying tags, the maintenance of cooperation was observed for a much wider region of the parameter space, reaching its peak at intermediate approval times and cost values above $c_{\mathrm{crit}}$. Our findings suggest that in an open system subject to immigration dynamics, high levels of social cooperation can be attained if a fraction of the population adopts the strategy with an egalitarian generosity directed towards both native and approved naturalized citizens, regardless of their actual origin. These findings also suggest that instead of relying upon arbitrarily fixed approval times, there is an optimal duration of the naturalization procedure from which the society as a whole can profit most. 

\end{abstract}


\end{frontmatter}

\section{\label{sec:level1}Introduction\protect}

The processes of migration and invasion dynamics, in a variety of their flavors and scales, are fundamental to our understanding of social, economic, ecological, and epidemiological phenomena~\cite{siminietal,song,huangzou,logan2002immigrant,stauffergerontol,franzoni,thober,sirkeci}. Largely due to the ongoing worldwide migration crisis, the studies of human mobility and immigration dynamics have garnered considerable attention in recent years across a wide variety of disciplines~\cite{saamorales,jiachristakis,lancet}, including physics~\cite{HMOrtmanns,zhixiwuPRE,xphan,vitanov,JEONG201947,WU2020124692}. 

Human mobile behaviors in general~\cite{siminietal,song,saamorales,xphan} and immigration flows in particular~\cite{HMOrtmanns,JEONG201947,WU2020124692}, 
can reach non-trivial levels of complexity, posing challenges to the efficiency of associated interventions and policy measures. This is especially the case if migrations occur unexpectedly, on massive scales, and in traditionally non-immigrant societies, often giving rise to the emergence of polarized attitudes~\cite{schahbasi2020}, segregation~\cite{schelling1971}, elevated intergroup tension~\cite{hewstoneetal}, and the enhanced expression of ethnocentric behaviors~\cite{HAmodel}.

From the view-point of evolutionary game theory one of the key issues in the study of open social systems subject to immigration dynamics is how to reach a persistence of sufficiently high levels of social cooperation when facing possible defective outcomes and costly integration processes. Models of the evolution of cooperation have traditionally ignored the topic of immigration, as they were largely based on the study of constant-sized populations of individuals inhabiting closed artificial systems that prevented an influx of any outsiders. A few notable exceptions have recently addressed a population's capacity to admit novel individuals that originated outside of the system~\cite{JEONG201947}, immigration effectiveness in dynamic networks~\cite{WU2020124692}, or the influence of immigrants' skills and diversity on the general welfare of the receiving society~\cite{lutzetal}. 

Unrealistically enough, these earlier models were limited to the study of phenotypically homogenous populations with fixed traits and were additionally constrained by employing only a few possible, unconditional strategies (i.e. pure cooperation and pure defection). Consequentially, most classical models of the evolution of cooperation did not investigate the influence of strategic diversity and phenotypic plasticity on the emerging differences in generosity among native and non-native individuals. However, under more realistic settings in phenotypically diverse societies, cooperation contingent on the opponent's phenotypic features or {\it tags} can naturally emerge, giving rise to a multitude of tag-mediated conditional strategies, even if such phenotypic labels are arbitrary and meaningless to the interacting parties~\cite{Efferson1844}. 

To address these realistic aspects, models of the evolution of tag-mediated cooperation~\cite{HAmodel,riolo2001,PhysRevE.68.046129,laird2012,HADZIBEGANOVIC20161,jansenbaalen,ramazi,hadzistauffhan,Jensen2019pre,hadzicuiwu,bravoyantseva,hadziliuli} have investigated the emergence of generosity in phenotypically diverse societies of artificial agents that were able to employ more than two pure strategies, and immigration of new agents in these models has often been considered as a distinct stage in the evolutionary process. One prominent example is the seminal work of Hammond and Axelrod (HA)~\cite{HAmodel}, addressing the ingroup-biased ethnocentric cooperation emerging in a phenotypically diverse population subject to immigration dynamics. However, the HA model and its subsequent extensions have studied the effects of immigration on the evolution of cooperative behavior in a rather less-systematic fashion, as they have used mostly constant immigration rates throughout the model simulations. 

One notable exception~\cite{HADZIBEGANOVIC20161} has recently examined the effects of variable immigration on the outbreak of tag-based cooperation, finding that the immigration rate can actually serve as a relatively good predictor of the density of ethnocentric cooperators under different mobility regimes. Thus, if various mobile behaviors are present within a population that is additionally open to immigration dynamics, the fraction of the dominant ethnocentric strategy in that population can be predicted from the immigration rate alone. These findings~\cite{HADZIBEGANOVIC20161}, and the results from subsequent studies~\cite{JEONG201947,WU2020124692,lutzetal}, have undoubtedly demonstrated that the choice of the specific immigration regime can critically alter 
the evolutionary outcomes in social dilemmas, which would otherwise remain undetected in closed systems simulated under constant population-size conditions. 

Importantly, none of these previous studies, neither tagless nor tag-based cooperation models, contained strategies conditional on immigrant-specific and  native-specific features, nor did they investigate the effects of the temporal heterogeneity of naturalization procedures of newcomer individuals on the evolution of global cooperation and competition among conditional and unconditional strategies in the increasingly diverse society that is open to immigration dynamics. Naturalization can be viewed both as a measure of inclusiveness and as a mechanism of social reproduction of a system~\cite{stauffergerontol}, whereby the resulting acquired citizenship~\cite{STEINHARDT2012813} plays a key role in the process of integration, sustenance of social cooperation, and in the formation of bonds between the native majority and immigrant minority groups~\cite{ferrera2019,politi2020}.

However, addressing these aspects would require not only an inclusion of novel strategies but also an implementation of dynamic features into the underlying model mechanisms. For instance, some features of newcomer agents may be subject to change as they adapt to the culture of the receiving society, yet most previous models of cooperation have studied the evolutionary dynamics under unchangeable, fixed trait conditions that once acquired were no longer able to evolve over an individual's lifetime.  

Combining the methods of statistical physics~\cite{PhysRevE.68.046129}, evolutionary game theory~\cite{anxo,adami}, and agent-based modeling and simulation~\cite{bonabeau}, we attempt to close these gaps by studying the coevolutionary dynamics of unconditional and multiple conditional strategies in an open multiagent system of native and non-native individuals with dynamically changing tags under continuous immigration dynamics, ageing, and heterogeneous naturalization approval times. More specifically, in our new model with six competing strategies, we introduce the concept of time-varying tags, such that the 'new' traits of immigrant agents can change to 'approved' features of naturalized citizen agents. 

In addition, instead of using only one parameter for determining the extinction probability of an agent, we introduce a higher level of granularity into the model's death mechanism: We distinguish between the initial extinction probability, and the 'posterior' extinction probability (after an agent has interacted with and imitated its neighbors), whereby the latter additionally depends on the relationship between an agent's payoff and the minimum requirements $M$ that must be satisfied. If an agent's payoff satisfies $M$, the extinction probability of an agent decreases by a given small value. Otherwise, the extinction probability grows by another small value $\alpha_{\mathrm{dis}}$; in our model, there is a specific value of this survival disadvantage $\alpha_{\mathrm{dis}}$ that optimally promotes cooperation, which depends on the naturalization approval time of new agents.

Besides the two unconditional and two conditional strategies that are typical for the standard HA model of ethnocentrism~\cite{HAmodel}, we further introduced two novel strategies in our tag-based cooperation model with time-varying tags: neophilia and sympolitic altrocentrism. Both of these strategies were introduced to reflect on the realistic emergence of more complex conditional behaviors that naturally evolve in multi-strategic and heterogeneous systems composed of different population subtypes (such as native and non-native individuals). 

As a result, our 6-strategy 2-tag model with dynamic tags and variable naturalization approval times comprises a highly heterogeneous population with a mixture of native and non-native individuals playing $12 \times 12$ spatial evolutionary games. In total, we studied two variants of this tag-based cooperation model, one with and another without the novel conditional strategies and time-varying tags, and we additionally compared these two model versions against the baseline model of the evolution of cooperation without tags and in the absence of any extinction probability and immigration dynamics.  

We found that altrocentric strategy, which cooperates with all approved citizens of the simulated society, not only wins over ingroup-biased 
ethnocentrism, but it also generates the highest levels of cooperation in our model under moderate naturalization approval times that persist even at remarkably large cooperation costs. On the other hand, we observe that ethnocentrism can only dominate under small costs and short approval times, whereas long approval periods lead to novel types of strategic coexistence, previously unreported in models of tag-based cooperation. 

Our findings suggest that without relaxing the naturalization procedures or introducing any specially permissive immigration policies, high levels of social cooperation can be attained if a fraction of the population adopts the altrocentric strategy with an egalitarian generosity directed towards both native and approved naturalized citizens, regardless of their actual origin. 

\section{\label{sec:level2}Model}

We consider a population of $N$ agents placed on the vertices of a regular square lattice with periodic boundary conditions. All agents are linked to their four nearest neighbors (Von Neumann neighborhood) with whom they can engage into pairwise interactions. During an interaction, agents play a Prisoner's Dilemma (PD) game with their neighbors. In this game, each agent chooses one of the two possible actions: cooperation or defection. Cooperation incurs a cost $c$ to the donating agent and yields a benefit $b$ to the recipient. The defection incurs no costs and yields no benefits. To satisfy the conditions of the PD game, $c$ should be positive and $b$ should be larger than $c$. In this paper, throughout our model simulations, we fix the value of $b$ to $b=1$ and we systematically vary the values of the cost $c$ to investigate its effects on cooperation.

Furthermore, there are two kinds of tags implemented in our model: new and approved. The new tags are time-varying, i.e. they can dynamically change due to the introduced approval time $\tau$: A new agent that invades the system from the outside always has the 'new' tag, but if this agent lives longer in the system than the designated approval time $\tau$, then this agent's tag will change its status to 'approved'.
We further consider six distinct strategies that can be adopted by the agents, four of which are conditional and the remaining two are unconditional. Specifically, the two pure strategies are the unconditional cooperation or altruism (A) and the unconditional defection or egoism (E). Two conditional strategies, intra-group ethnocentrism (I) and extra-group cosmopolitanism (O), are the same as in most previous models of tag-based cooperation~\cite{HAmodel,laird2012,HADZIBEGANOVIC20161,hadzistauffhan,hadzicuiwu}. Ethnocentric (I) individuals cooperate only with the opponents who share the identical tag; otherwise, they always defect with others who display a different tag. Cosmopolitan out-group cooperators (O), on the other hand, always cooperate with the opponents whose tags are distinct from their own; otherwise, they always decline to cooperate. The remaining conditional strategies in our present model, neophilia (N) and sympolitic altrocentrism (S), are the two newly proposed strategies. Neophilic agents cooperate only with individuals who display the new tag; otherwise, they always defect. Sympolitic (S) altrocentrists (from the Greek word $\Sigma \upsilon \mu \pi \textomikron \lambda \acute{\iota} \tau \eta \varsigma $, meaning 'fellow citizen') cooperate only with others who carry the approved 'citizen' tag; otherwise, they always defect (hereafter, and for simplicity, we denote this strategy as altrocentric or S-strategy). In addition, the term 'sympolitic' is chosen for another reason, namely, in the context of citizenship: As opposed to the 'isopolitical' view, the concept of citizenship in nation states is typically sympolitical, i.e. its contents, scope, and allocation criteria are designed by central authorities via democratic procedures~\cite{ferrera2019}.

Thus, the altrocentric S-agents in our model are benevolent to all agents displaying the approved 'citizen' tag, regardless of whether such agents have only recently become 'naturalized' or whether they are 'old' native citizens, in the sense that many generations of their ancestors have also been native agents. Generally, altrocentrism~\cite{stagner} represents a somewhat more socialized ability to understand other individuals by means of viewing things and events as they appear to them, from their own viewpoints. As a result, altrocentric agents perceive themselves (and others) not as separate entities but instead as integral parts of the whole social system or the whole community. On the other hand, the malevolence of altrocentric agents towards individuals with 'new' tags was implemented in our model mainly to reflect on the realistic limitations that are imposed upon newcomers in their receiving societies prior to accomplishing specific approval procedures, such as residence and work permits or the citizenship status, without which they cannot fully take part in the host society. 

\begin{figure}
\centering
\includegraphics[angle=270,width=1.1\columnwidth, angle=90]{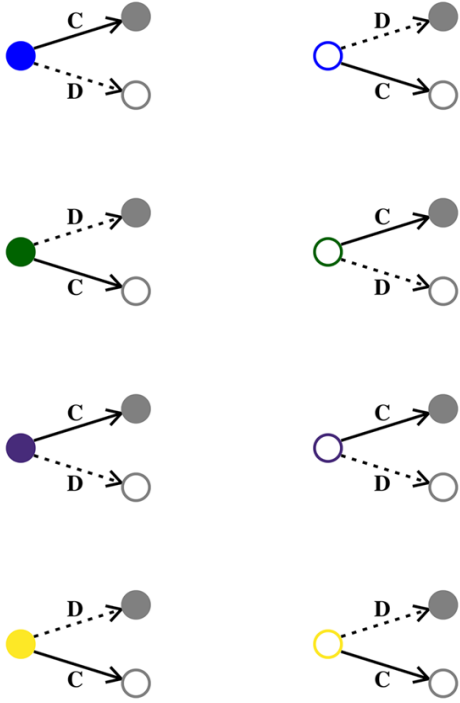}
\caption{Conditional strategist behaviors against opponents with 'approved' and 'new' tags in the 6-strategy 2-tag model of cooperation with time-varying tags. Agents with blue color are ethnocentric, green are cosmopolitans, purple agents are altrocentric, and yellow are neophilic. Solid and dashed arrows stand for cooperation (C) and defection (D), respectively. Full and open circles represent 'approved' and 'new' agents. Gray color of the opponent co-players stands for any of the six possible strategies. }
\label{stratfig}
\end{figure}

In summary, our two novel, conditional strategies N and S, are in a sense discriminating~\cite{jensen2019imitating} strategies as they focus only on a single feature of agents, and can therefore be viewed as special cases of the ingroup-biased ethnocentric (I) and the out-group biased cosmopolitan (O) strategies. The total of six strategies employed in our model and the associated resulting behaviors (cooperation or defection) given the displayed tags (new vs. approved) of interacting parties, are further summarized in Table~\ref{tab1} and in Fig.~\ref{stratfig}. In addition, the complete $12 \times 12$ matrix of exact payoffs resulting from the PD-based interactions of agents with six different strategies (A, E, I, O, N, S) and two different tags ('new' and 'approved') in our model is shown in equations (A.1) and (A.2) of the Appendix. 

\begin{table}
\caption{Behaviors of strategists given their own tags and the tags of the opponents.}\label{tab1}
\begin{tabular}{ll|ll|}
\cline{3-4}
\multicolumn{2}{l|}{}                                             & New       & Approved  \\ \hline
\multicolumn{1}{|l}{\multirow{2}{*}{Altruism (A)}}             & New      & Cooperate & Cooperate \\
\multicolumn{1}{|l}{}                                  & Approved & Cooperate & Cooperate \\ \hline
\multicolumn{1}{|l}{\multirow{2}{*}{Egoism (E)}}             & New      & Defect    & Defect    \\
\multicolumn{1}{|l}{}                                  & Approved & Defect    & Defect    \\ \hline
\multicolumn{1}{|l}{\multirow{2}{*}{Ethnocentrism (I)}}    & New      & Cooperate & Defect    \\
\multicolumn{1}{|l}{}                                  & Approved & Defect    & Cooperate \\ \hline
\multicolumn{1}{|l}{\multirow{2}{*}{Cosmopolitanism (O)}}  & New      & Defect    & Cooperate \\
\multicolumn{1}{|l}{}                                  & Approved & Cooperate & Defect    \\ \hline
\multicolumn{1}{|l}{\multirow{2}{*}{Altrocentrism (S)}}        & New      & Defect    & Cooperate \\
\multicolumn{1}{|l}{}                                  & Approved & Defect    & Cooperate \\ \hline
\multicolumn{1}{|l}{\multirow{2}{*}{Neophilia (N)}}        & New      & Cooperate & Defect    \\
\multicolumn{1}{|l}{}                                  & Approved & Cooperate & Defect    \\ \hline
\end{tabular}
\end{table}

Each simulation in our present study begins with a network comprised of $N$ agents who are randomly assigned their strategy and the tag without any bias. Based on the assigned strategy and the tag, each agent plays the PD game with all of its neighbors and gathers the resulting payoff. Next, by means of an imitation process, all agents have a chance to change their strategies at the same time synchronously. Specifically, an agent $i$ chooses one of its neighbors $j$ randomly and imitates its strategy according to the imitation probability
\begin{eqnarray}
 P_{i \rightarrow j} = \begin{cases} \frac{\pi_j-\pi_i}{(b + c) \times \mathrm{max}\{k_i,k_j\}} & \text{if}\; \pi_j > \pi_i  \\ 0 & \text{if}\; \pi_j \leq \pi_i \end{cases}, \label{EQ1}
\end{eqnarray}
where $\pi_i$ and $k_i$ are the payoff and the degree (the number of neighbors) of the agent $i$, respectively~\cite{santos2006evolutionary,gomez2007dynamical}. In this work, $k_i=4$ for all agents.

A fitter agent has a higher chance to survive~\cite{nowak2004evolutionary,nowak2006five}, and so we adopt the minimum requirements (aspiration level, $M$) and the extinction probability ($\alpha$)~\cite{Macy7229,JEONG201947,LIU2019247,Zhang_2019,WU2020124692}. 
The extinction probability indicates the probability that an agent disappears from the network at one time step. When an agent appears in the network, it has the initial extinction probability $\alpha_{\mathrm{ini}}=0.0005$. After the imitation process, however, each agent updates its extinction probability: if the payoff $\pi_i$ of the agent $i$ satisfies the minimum requirements ($\pi_i \geq M$), the extinction probability decreases by $\alpha_{\mathrm{sat}}=0.0005$; otherwise, it increases by $\alpha_{\mathrm{dis}}$. In this paper, $\alpha_{\mathrm{dis}}$ is set to $0.0008$ except for Fig.~\ref{fig7}, where the effect of $\alpha_{\mathrm{dis}}$ on cooperation under different approval time $\tau$ is systematically studied. 

We set the minimum extinction probability to $\alpha_{\mathrm{min}}=0.0001$ so that agents cannot live forever. Then, the extinction probability $\alpha_i(t)$ of agent $i$ at time step $t$ is given by
\begin{eqnarray}
\alpha_i (t) = \begin{cases} \alpha_i (t-1) - \alpha_{\mathrm{sat}} & \pi_i \geq M\\
\mathrm{max}\!\left\{\alpha_i (t-1) + \alpha_{\mathrm{dis}}, ~\alpha_{\mathrm{min}} \right\} & \pi_i < M \end{cases} . \label{EQ2}
\end{eqnarray}
After the update of the extinction probability, each agent has a chance to die according to the underlying extinction probability. 
This series of procedures in our model is called the extinction process.

\begin{figure*}
\vspace{-1.15cm}
\centering
\includegraphics[angle=270,width=2\columnwidth]{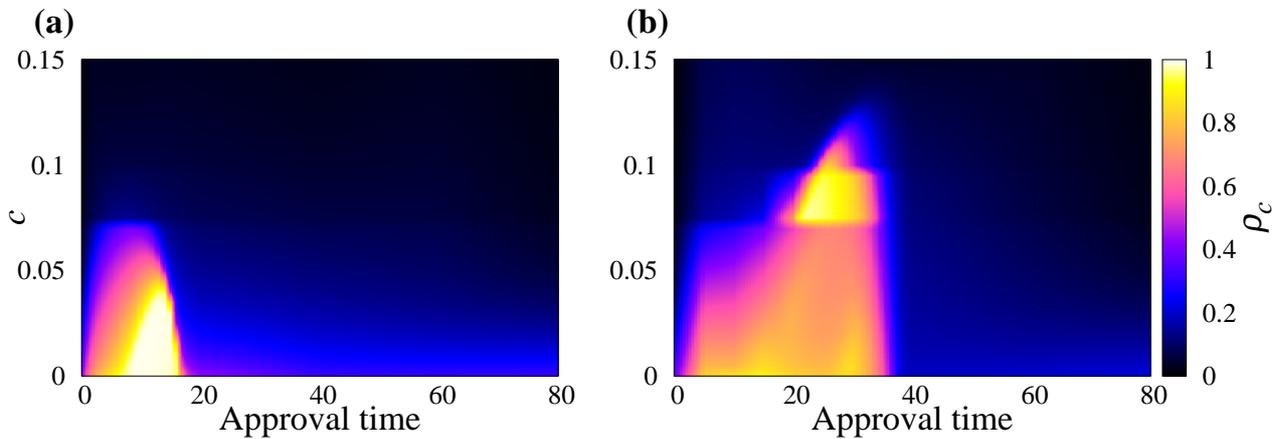}
\caption{Cooperation level $\rho_c$ as a function of approval time $\tau$ and cost $c$ for the standard 4-strategy model of tag-based cooperation with 
fixed tags (a), and the generalized 6-strategy model of tag-based cooperation with novel strategies and time-varying tags (b). 
The remaining model parameter values are listed in Methods. }
\label{fig1}
\end{figure*}

\begin{figure*}
\vspace{-0.691cm}
\centering
\includegraphics[angle=270,width=2\columnwidth]{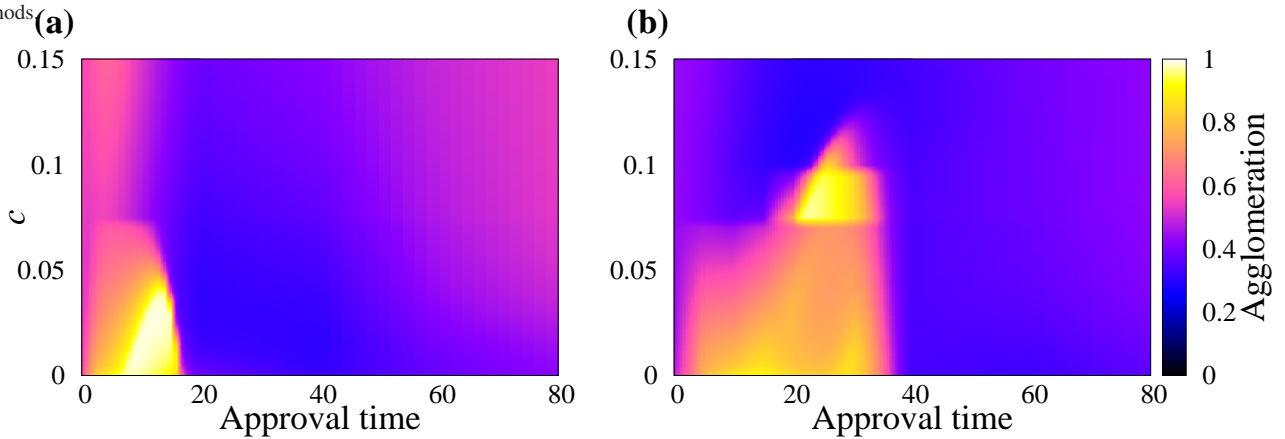}
\caption{Agglomeration degree as a function of approval time $\tau$ and cost $c$. The standard 4-strategy model with 
fixed tags (a), and the generalized 6-strategy model with time-varying tags (b). The other model parameters were set to their baseline values as listed in Methods.}\label{fig2}
\end{figure*}

As a result of the extinction process, some agents die out, leaving their network nodes empty and available for new agents. All vacant nodes are then occupied by new agents with new tags and randomly assigned strategies. This is the immigration process. Following the immigration process, the new immigrant agent that invaded the system from the outside is assigned the age of 1, and the age of all other agents (already inhabiting the system) is increased by one. Thus, more precisely, our 'ageing' process actually describes the time spent living within the system, rather than the actual age of individuals. Finally, each agent is subjected to an approval process: It is first checked if there are any agents in the system with the 'new' tags. If yes, it is then checked for each individual 'new' agent if their time spent in the system is greater than the designated approval time $\tau$; if yes, the 'new' tags are then changed to 'approved'. 

One iteration (time step) of our simulated evolutionary process is thus composed of a total of six different evolutionary stages: interaction with payoff calculation, imitation (strategy-reproduction), extinction, immigration, ageing, and approval process. 

In this paper, individual simulation results were obtained by averaging over the last $5000$ generations taken after a transient period (at least $20000$ generations), and the outcomes of $30$ independent simulation runs with different initial random number seeds were then averaged for the final results.

\section{\label{sec:level3}Results and Discussion}

Unless otherwise specified, all results in the present study were obtained from simulations performed on $200 \times 200$ regular square lattices with periodic boundary conditions. Since the baseline value of minimum requirements in our model was set to $M= 2.7$, the agents needed at least three cooperative neighbors to satisfy the minimum requirements condition. We note that our results did not change qualitatively if the value of the minimum requirements was between 2 and 3 (in Fig.~6 and Section~3.3, we present a systematic analysis and a discussion of the effects of minimum requirements, and their combined effects with approval time on the maintenance of cooperation in our model).

\begin{figure*}[t]
\centering
\includegraphics[angle=270,width=2\columnwidth]{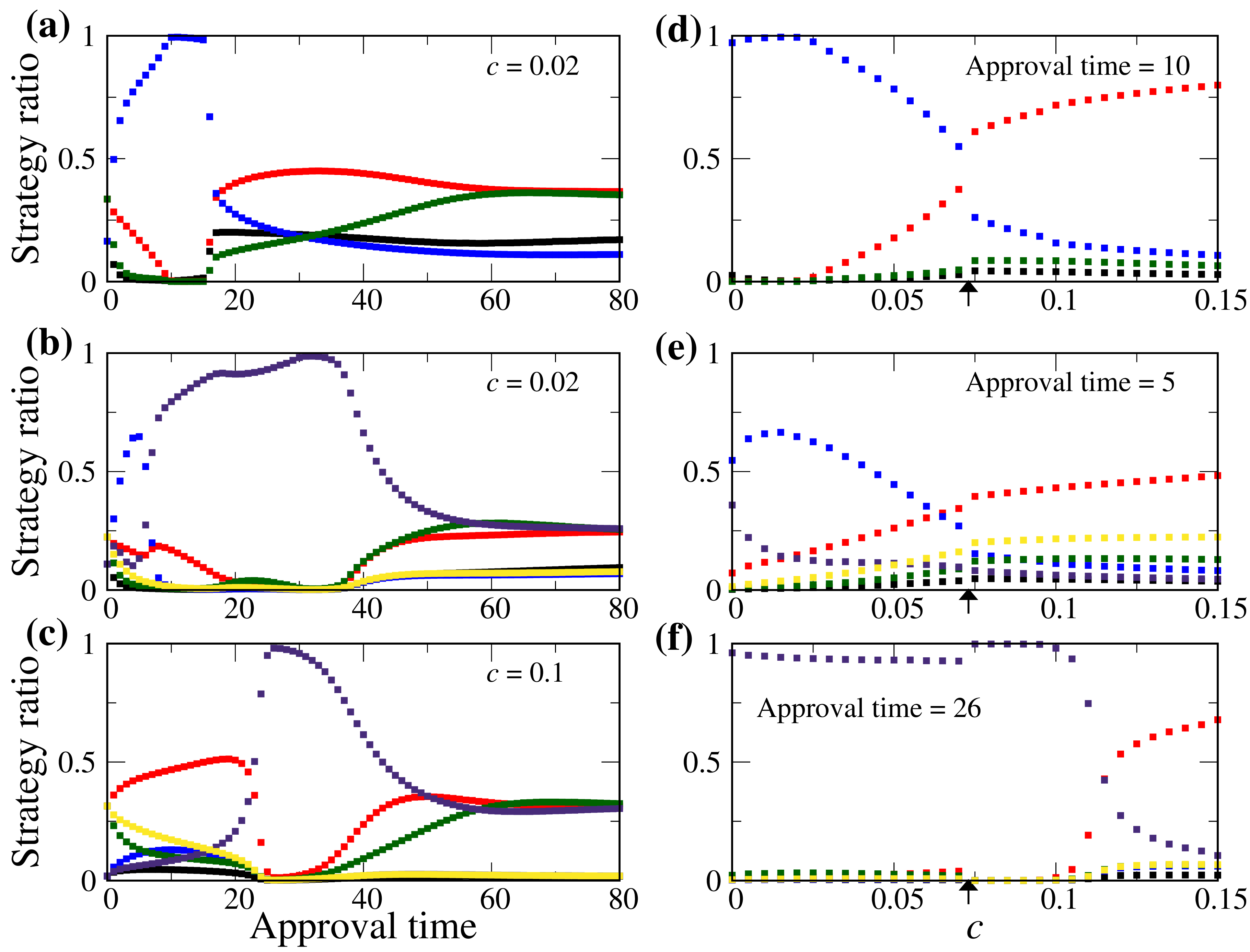}
\caption{Strategy ratios as a function of different approval times $\tau$ for a given fixed cost (left column), and as a function of varied cost $c$ for a given fixed approval time (right column). The other model parameters were set to their baseline values as listed in Methods. (a) The standard 4-strategy 2-tag model of cooperation with fixed tags. (b) The generalized 6-strategy 2-tag model of cooperation with novel discriminating strategies and time-varying tags. (c) The generalized 6-strategy 2-tag model at a higher cost condition $c=0.1$. (d) The standard 4-strategy 2-tag model with a fixed approval time. (e) The generalized 6-strategy 2-tag model with a fixed approval time. (f) The generalized 6-strategy 2-tag model in a longer approval time condition. Black squares stand for altruism, red for egoism, blue for ethnocentrism, green for cosmopolitanism, purple for altrocentrism, and yellow represents neophilia.}
\label{fig3}
\end{figure*}

\begin{figure*}[htb]
\centering
\includegraphics[angle=270,width=2\columnwidth]{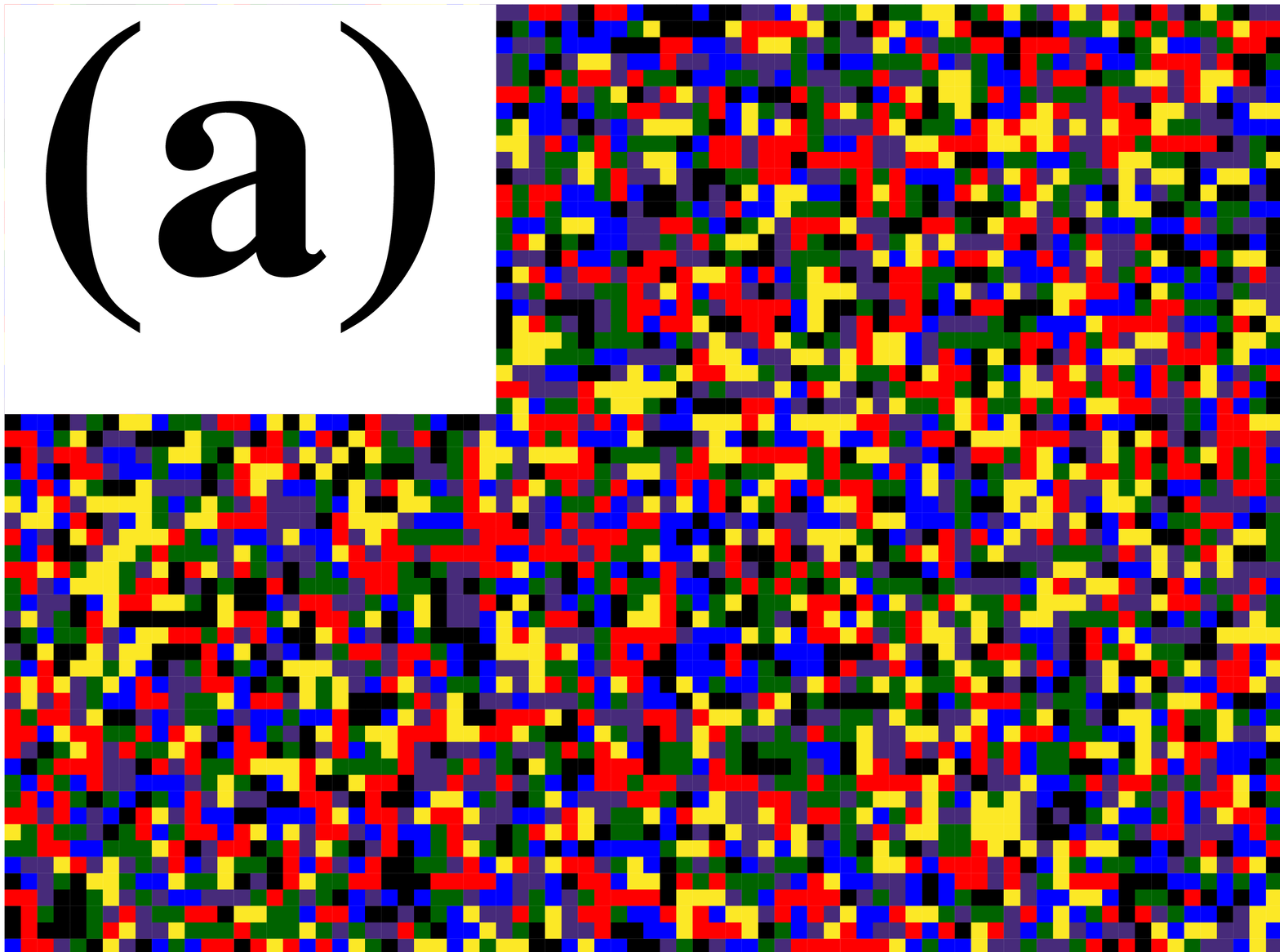}
\caption{Typical two-dimensional color snapshots of the evolutionary dynamics of strategies in our 6-strategy 2-tag model of cooperation with time-varying tags. All snapshots were obtained with the cost value $c=0.1$, the approval time $\tau=26$, and $\alpha_{\mathrm{dis}}=0.0008$. The system size was $150 \times 150$ and the other model parameters were set to their baseline values as detailed in Methods. Black color in the snapshots stands for the altruistic strategy, red for egoists, blue for ethnocentrism, green for cosmopolitanism, purple for altrocentrism, and yellow represents neophilia. Tags are not shown in this figure. (a) The initial state $t=0$ with randomly distributed strategies. (b) Strategy distribution after $t=100$ generations. (c) After $t=300$ generations. (d) After $t=1000$ generations. (e) After $t=3000$ generations. (f) After $t=30000$ generations (the equilibrium state).}\label{fig4}
\end{figure*}

\subsection{\label{sec:level4}Cooperation level and agglomeration with and without the novel strategies}
 
We first studied the cooperation level $\rho_c$ as a function of the approval time $\tau$ and cost $c$. Since we employed a mix of unconditional and conditional strategies with time-varying tags, conditional strategists in our model were both cooperators and defectors, as they cooperated with some agents but defected with others. Therefore, as a more meaningful measure of cooperative behavior in our model, we calculated the level of cooperation $\rho_c$, which we here 
define as the ratio of cooperative edges such that both end-nodes' co-players $i$ and $j$ connected by the edge $\epsilon_{ij}$ cooperated with one another.

When the approval time is zero in our model, all individuals will have the approved tag, and so the agents with strategies altruism (A), ethnocentrism (I), or altrocentrism (S) will always cooperate, whereas the agents with any of the three other strategies will always defect. Thus, at approval time zero, we have a special case which is technically equivalent to the tagless model with unconditional cooperation and unconditional defection. In this special case, cooperative agents can hardly survive as they are continuously exploited by defective individuals; therefore, the cooperation level in this particular condition is markedly low. Similarly, when the approval time is very long, the majority of agents will not accomplish the approval procedure and will therefore have the new tag. For the same reason as before, we have another special case equivalent to the tagless model, and the overall cooperation level remains low. Thus, most agents do not satisfy the minimum requirements under this scenario and consequentially they die out before even reaching an approval. The evolutionary outcomes that emerge under these two limiting cases are verified in Fig.~\ref{fig1}.

Remarkably, for an intermediate approval time $\tau$, the time-varying tag system becomes effective and the two discriminating novel strategies start to play nontrivial roles in the model. As can be observed in Fig.~\ref{fig1}, the models with and without the two novel strategies can both exhibit very high levels of cooperation under adequate approval times. However, relative to the standard 4-strategy model of the evolution of tag-based cooperation, our correspoding 6-strategy model generalization with the two novel discriminating strategies covers a substantially larger area of the parameter space for which high cooperation levels can be attained. In addition, and somewhat counter-intuitively, there is a region of the parameter space for which the level of generosity remains strikingly high in spite of a relatively large cost of cooperation ($0.073\lesssim c \lesssim 0.1$).

In Fig. 2 we see that cooperation depends on both $\tau$ and $c$. In the standard 4-strategy model, cooperation is well-maintained only at lower values of both $\tau$ and $c$. Above the critical point $c_{\mathrm{crit}}$, cooperation transitions abruptly to a state of dominant defection, and already in the lower range of intermediate approval times, like $\tau < 20$, cooperation becomes vanishingly weak without recovery at longer approval times. Remarkably, particularly high levels of cooperation in our 6-strategy model with time-varying tags are maintained at high cost values $c_{\mathrm{crit}} < c < 0.1$ but only at intermediate (such as $20 < \tau < 39$) and not at low or high values of $\tau$. Thus, both $c$ and $\tau$ influence cooperation, and as we can see more clearly in Fig. 4, there is an interaction effect between $\tau$ and $c$ on the cooperation level both in our 6-strategy model with time-varying tags and in the standard 4-strategy model with fixed tags.

When playing tagless but spatial PD games, agents' survival critically depends on their ability to assort with cooperative others. In tag-based cooperation models, spatial interactions are not a necessary prerequisite for the emergence of cooperation, as cooperation via tags is viable also in aspatial games~\cite{hadzicuiwu}. However, spatial assortment of agents with ingroup-biased strategies and the resulting emergence of ethnocentric clusters can significantly enhance cooperation levels in structured populations~\cite{hadzicuiwu,hadziliuli}. Thus, cooperation can be markedly elevated when agents adjoin mutually cooperative co-players of the same strategy and tag~\cite{xia2012effect,Roca11370}.

The degree of this adjoining processes and the associated differences across established clusters can be represented via agglomeration \cite{Roca11370}, which is defined as
\begin{eqnarray}
\text{Agglomeration} = \sum_{i}^{N} \sum_{j \in N_i} \frac{\sigma_{ij}}{k_i}. \label{EQ3}
\end{eqnarray}
where $N_i$ and $k_i$ represent the neighbors and the degree of the agent $i$, respectively. The $\sigma_{ij}=1$ when the strategy and the tag of the agent $j$ are the same as the strategy and the tag of agent $i$; otherwise, we have $\sigma_{ij}=0$. Fig.~\ref{fig2} shows the resulting agglomeration as a function of approval time and cost $c$ in the standard 4-strategy model and in our 6-strategy model with time-varying tags. Comparing Figs.~\ref{fig1} and~\ref{fig2} reveals that agglomeration is highly correlated with cooperation level in our model. This suggests that high cooperation is most likely induced by solid clusters of players employing the same strategy and tag. Furthermore, from Fig.~7 we can surmise that the resulting cooperative cluster is composed largely of agents carrying the approved tag.

However, we note that unlike the cooperation level, the degree of agglomeration is still relatively high for both very short and very long approval times, 
and is the highest in the region of the parameter space corresponding to the high cooperation levels. Notably, the degree of agglomeration for very short and very long approval times is higher in the 4-strategy model variant with fixed tags than in our present, generalized 6-strategy model version with two novel discriminating strategies and time-varying tags (Fig.~\ref{fig2}). We tentatively conclude that this phenomenon is largely due to the greater strategic diversity in the latter case, where cluster formation is more challenging in the presence of 
multiple competing strategies.

Even if an agent satisfies the minimum requirements for a long approval time, it still has a chance to die because of the non-zero minimum extinction probability. However, we see that in accordance with Figs.~\ref{fig1} and~\ref{fig2}, both high cooperation and high agglomeration remain stable across a range of values of $\tau$ and $c$ despite this permanent extinction vulnerability.  

\subsection{\label{sec:level9}Cooperation-promoting strategies}

The results presented in the previous subsection revealed that clusters consisting of the same strategy and the same tag coincided with the area of particularly high cooperation. To identify the underlying strategy responsible for the formation of these clusters, we systematically assessed the ratios of strategies as a function of approval time $\tau$ and cost $c$. In the standard model without novel strategies, the ethnocentric (I) strategy dominated over all others in the high cooperation-level area (see Figs.~\ref{fig3}(a) and~\ref{fig3}(d)). Cooperative clusters were thus largely composed of ethnocentric agents, and in this model of tag-based cooperation without novel strategies but with an implemented approval mechanism, ethnocentrism was the only strategy that promoted high levels of cooperation, especially at lower costs and shorter approval times. As the approval time and the cost of cooperation increased in this standard model version, defection took over the population, even though at very long approval times other competing strategies were not fully suppressed. Instead, due to random assignments of strategies to the increasingly invading 'new' agents, cosmopolitans started to coexist with egoists at roughly equal levels. 

Our generalized 6-strategy model with time-varying tags exhibited a strikingly different behavior. Here, Figs.~\ref{fig3}(b) and~\ref{fig3}(e) show that the dominance of a given strategy strongly depended on the underlying approval time: Sympolitic altrocentrism consistently outweighed all other strategies except for very short and very long approval times, where either ethnocentrism took over the population (at fast approvals) or the three strategies coexisted in time (at delayed approvals). However, this prevalence of ethnocentrism at very short approval times (e.g. at $\tau=5$) was observed only at sufficiently low but not at higher values of the cost $c$. These findings thus indicate the existence of an interaction effect between $\tau$ and $c$ on the density of ethnocentric cooperators both in our 6-strategy model with time-varying tags, and in the standard 4-strategy model with fixed tags. Furthermore, we see a similar interaction 
effect on the density of altrocentric agents that dominate at a range of intermediate approval times and cost values, but not when the costs exceed the value $c > 0.1$, after which the egoist strategy again takes over.
 
Remarkably, at higher cost values such as $c=0.1$ (Fig.~\ref{fig3}(c)), elevated cooperation levels were generated only by the altrocentric (S) strategy,  
as the costs here exceeded the limit under which the ethnocentrism can promote cooperation. We also see that the coexistence of the three strategies (S, C, and E), previously observed at very long approval times and a lower cost $c=0.02$, is also given at substantially higher cost values such as $c=0.1$. Unlike at moderate approval times where cosmopolitan strategy was vanishingly weak, extra-group E cooperators can rise to nonnegligible levels at larger values of $\tau$ and then stably coexist in an arms race with S and C strategists at very long approval times, ultimately reaching substantially higher levels. This finding is interesting, because stable levels of heterophilic cosmopolitan strategy around or above 25\% percent of the population were previously rarely observed in tag-based cooperation models~\cite{laird2012,hadziliuli}.

In summary, the main strategy promoting cooperation in our novel 6-strategy model with time-varying tags is sympolitic altrocentrism that is marked by egalitarian cooperative attitudes towards all approved fellow 'citizens', whereas the exclusively ingroup-biased ethnocentrism is effective only in a narrow area of the parameter space characterized by short approval times and very low costs.

Interestingly, as shown in Figs.~\ref{fig1} and \ref{fig3}, there is a critical value of cost $c_{\mathrm{crit}}\cong0.073$, at which strategy ratios and cooperation levels change discontinuously (the position of $c_{\mathrm{crit}}$ is highlighted by the arrows on the abscissa in Figs.~\ref{fig3}(d)-(f)).
We can see in Fig.~\ref{fig1}(a) that regardless of the underlying approval time, there is an abrupt transition to a dominant defective state that occurs above this critical cost $c_{\mathrm{crit}}$ in the standard 4-strategy model of tag-based cooperation with fixed tags. Thus, the equilibrium density of ethnocentric individuals and the overall cooperation level drop abruptly as cost increases at this critical point (see Figs.~\ref{fig3}).

It is worth noticing here that the value of $c_{\mathrm{crit}}$ was found to coincide with the cost at which cooperators are completely eliminated by defectors in the much simpler model without tags and in the absence of any extinction and immigration processes (not shown). We therefore infer that the existence of the critical cost value in our model is related only to the payoff matrix and the underlying network structure.

In our generalized 6-strategy model with time-varying tags, a similar behavior can be observed at short approval times. However, at intermediate approval times, cooperation increases sharply and reaches its peak as the cost exceeds the critical value $c_{\mathrm{crit}}$, and then gradually decreases and changes into the state of dominant defection after $c > 0.1$. This counter-intuitive boost of cooperative behavior observed above $c_{\mathrm{crit}}$ was driven 
by a significant rise in the frequency of altrocentric cooperators, which were the only type of strategists in our model that could thrive even in the presence of very high cooperation costs.  

\begin{figure}[htb]
\centering
\includegraphics[angle=270,width=1\columnwidth]{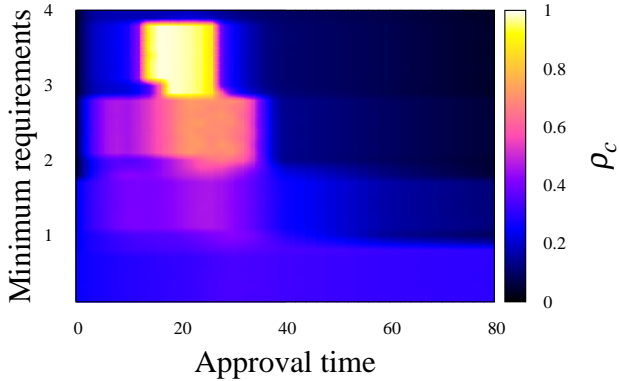}
\caption{Cooperation level $\rho_c$ as a function of approval time $\tau$ and minimum requirements $M$ at a fixed cost $c=0.05$ for our generalized 6-strategy 2-tag model with time-varying tags. The other model parameters were set to their baseline values as detailed in Methods.}
\label{appendixfig1}
\end{figure}

\begin{figure}[tb]
\hspace{-0.39cm}
\centering
\includegraphics[angle=270,width=1.041\columnwidth]{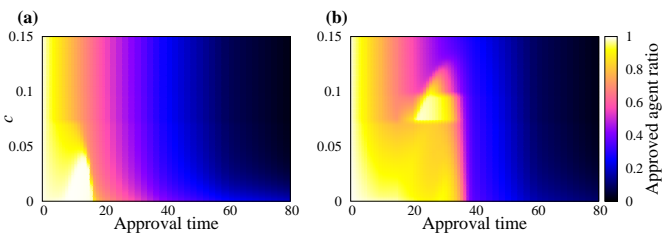}
\caption{The ratio of approved agents as a function of $\tau$ and $c$ for the standard 4-strategy 2-tag model with fixed tags (a) and our generalized 6-strategy 2-tag model with time-varying tags (b). The other model parameters were set to their baseline values as detailed in Methods.}
\label{appendixfig2}
\end{figure}

\subsection{\label{sec:level5}Cluster growth analysis in a cooperative network}

Fig.~\ref{fig4} shows the evolution of the model behavior as it settles into different states over time. It captures the growth process of cooperative 
clusters dominated by the altrocentric strategy (at cost $c=0.1$ and approval time is $\tau=26$):
(a) Initially, each agent has a randomly assigned strategy;
(b) Some altrocentric (purple) but also small neophilic (yellow) and ethnocentric (blue) clusters start to form, while the majority of agents still has the egoistic strategy (red);
(c) Clusters of altrocentric agents visibly grow in size;
(d) Altrocentric clusters expand further and become interconnected, starting to form one giant cluster;
(e) Altrocentrism becomes the dominant strategy;
(f) After sufficiently many generations, most agents have the altrocentric strategy, and all other competing strategies are drowning in the sea of altrocentrism. 
In cases where the population is dominated by the ingroup-biased ethnocentrism, i.e. at very short approval times $\tau$ and cost values below $c_{\mathrm{crit}}$, ethnocentric clusters exhibit similar growth patterns as observed in the expansion of altrocentric agents in Fig.~\ref{fig4}. 

Fig.~\ref{appendixfig1} shows the the cooperation level $\rho_c$ as a function of minimum requirements $M$ and approval time $\tau$. We see that there is a region of the parameter space characterized by very high levels of cooperation, and that within this region, the actual value of $M$ hardly affects $\rho_c$. 
Although in most of this work we fixed the minimum requirements to $M=2.7$, the resulting outcomes remained qualitatively the same even if we varied $M$ within this region of the highest cooperation. In Fig.~\ref{appendixfig2} we show the approved agent ratio as a function of $\tau$ and $c$. Obviously, the approved agent ratio is very high at short approval times since the new agents are here approved rather quickly. However, besides this trivial case, the approved agent ratio is also remarkably elevated in the range of values of $\tau$ and $c$ for which the overall cooperation level is also high, which further implies that high cooperation levels in our model are typically reached via interactions among approved agents.

\begin{figure}[tb]
\centering
\includegraphics[angle=270,width=1\columnwidth]{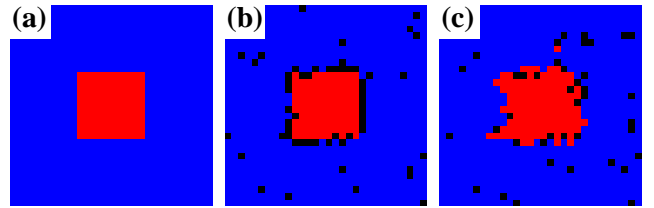}
\caption{Growth of the altrocentric cluster with 'approved' agents in our 6-strategy 2-tag model of cooperation with time-varying tags. The lattice 
size was $30 \times 30$, and the values of the remaining parameters were the same as described in the caption of Fig.~\ref{fig5}. Red means an 'approved' agent with altrocentric strategy, and black is a 'new' agent with altrocentric strategy. Blue color represents agents with all other kinds of strategies and tags.   (a) The initial state with centrally positioned altrocentric agents with 'approved' tags. (b) After $t=20$ generations. (c) After $t=30$ generations.}\label{fig5}
\end{figure}

Details of the spreading process of altrocentric clusters are further depicted in Fig.~\ref{fig5}: 
(a) A centrally placed cluster composed of 'approved' agents with altrocentric strategy (red) is surrounded by all other kinds of agents displaying different tags (blue);
(b) Around the 'approved' cluster of altrocentric individuals (red), 'new' agents (black) start to aggregate, imitating the altrocentric strategy as agents inside the cluster gain relatively higher payoffs;
(c) After sufficiently many generations, 'new' agents have accomplished the approval process and join the cluster with their newly 'approved' tags.
As this spreading process is iterated, the altrocentric cluster with 'approved' individuals expands further forming one giant cluster at the expense of all other competitors.

\begin{figure}[tb]
\centering
\includegraphics[angle=270,width=1\columnwidth]{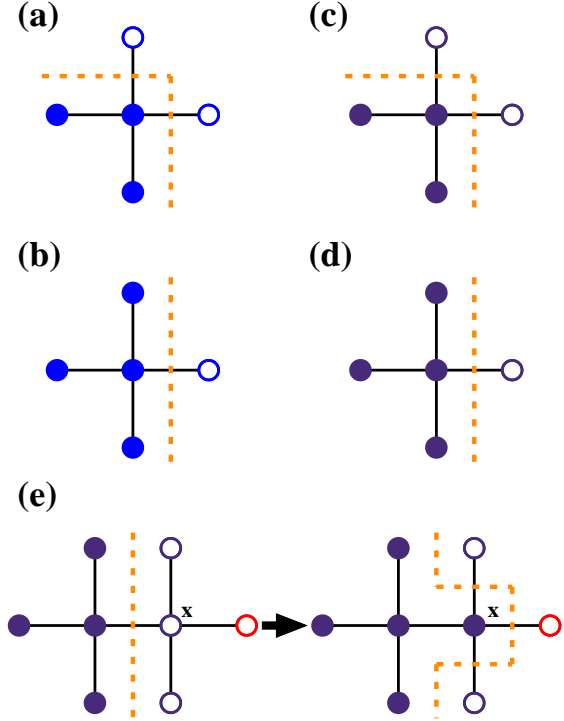}
\caption{Different interaction scenarios at the boundary of cooperative clusters. Filled and empty circles represent agents with 'approved' and 'new' tags, respectively. The dashed orange lines are the cluster boundaries. Blue is for ethnocentrism, purple is for altrocentrism, and red is for egoism. An example of a spreading process of a small altrocentric cluster is depicted in (e), where an altrocentric 'approved' agent expands at the cost of a 'new' altrocentric agent.}\label{fig6}
\end{figure}

Thus, the two key ingredients for attaining a highly cooperative system are the formation of cooperative clusters and their sustainable growth process. Notably, ethnocentric clusters have an initial reproductive advantage over altrocentric and other types of clusters, because adjacent ethnocentric co-players sharing either 'approved' or 'new' tags will always cooperate with one another, and as 'new' agents become 'approved', they can continue this mutualism and establish stable ethnocentric cluster formations that will not be affected so heavily by the tag-changing dynamics. On the other hand, since altrocentric agents cooperate only with 'approved' co-players, the emergence of stable altrocentric clusters strongly depends upon adequate approval times. As we have seen from our simulation experiments, the cluster formation of altrocentric players under optimal (intermediate) approval times can beat the clustering efficiency of their ethnocentric competitors, as only 'approved' ethnocentrics can exploit altrocentric cooperators, whereas both 'approved' and 'new' altrocentrics can benefit from their ethnocentric opponents. 

To gain a better understanding into the prevalence of cooperators and the dominance of the altrocentric strategy that is largely associated with it, we examined in Fig.~\ref{fig6} the competition between ethnocentric (I) and altrocentric (S) agents at the boundary of a cooperative cluster. In this example, we assumed for simplicity that the cluster is surrounded by 'new' agents employing the same strategy as the cluster members. Then the payoffs of the ethnocentric agents at the corner and the edge of the cluster are $2(1-c)$ and $3(1-c)$, respectively. However, these payoffs are lower than those of the altrocentric agents, which are $(4-2c)$ at the corner and $(4-3c)$ at the edge. 

Since for the most simulations conducted in our present paper we assumed that the value of the minimum requirements was $M=2.7$, the ethnocentric agent at the corner cannot satisfy these minimum requirements and thus the extinction probability of this agent consequentially increases. As a result, this ethnocentric agent would die out with high probability before any 'new' agents surrounding the cluster become 'approved', unless the approval time is short enough. Through the loss of the corner agent, the ethnocentric cluster would shrink recursively. We note here once again that there were no qualitative differences whatsoever in our simulation outcomes when the value of the minimum requirements was changed to any other value between 2 and 3.

For the ethnocentric cluster to grow in size, new agents surrounding the cluster have to be approved and need to enter into the cluster before the death of the corner agent. As for the ethnocentric agent at the edge, the payoff $3(1-c)$ is larger than the value of the minimum requirements $M$, but only for a sufficiently small cost. Therefore, ethnocentric agents at the edge of the cluster become unstable for relatively large cost values. This is the main reason why ethnocentrism retains its dominance only at short approval times and low cooperation costs. 

On the other hand, the 'approved' altrocentric agents at the boundary of a cooperative cluster obtain the benefits from all 'new' agents with the altrocentric strategy, thereby surviving even until the 'new' agents are 'approved' and merged into the growing cluster. Thus, 'new' agents surrounding the cluster composed of 'approved' altrocentric individuals can become 'approved' and join the cluster if the underlying approval time is sufficiently long. To understand how the cluster grows in size, we need to know the payoff of the agent \textbf{x} across the cluster boundary before and after its approval (Fig.~\ref{fig6}(e)).

In Fig.~\ref{fig6}(e) we see an example of how a 'new' agent can enter the formed altrocentric cluster. Before the approval, the payoff of the agent \textbf{x} is $-c$, which is less than the minimum requirements $M$. But if the agent \textbf{x} survives until its 'new' tag becomes 'approved', the resulting payoff 
then increases to $(3-c)$, which satisfies the minimum requirements for any values of the cost considered in this present work (see the right panel of Fig.~\ref{fig6}(e)). Since the payoff of the agent \textbf{x} is higher than that of its neighbors, the agent \textbf{x} then manages to maintain 
its strategy and the neighbors start to imitate the strategy of \textbf{x} with high probability. As this process occurs repeatedly throughout the simulation, the size of the established altrocentric cluster grows. If the growth of such cooperative clusters (ethnocentric or altrocentric) passes a tipping point, they become irreversibly interconnected forming one giant cooperative cluster.

\subsection{\label{sec:level7}The relationship between the approval time and the survival disadvantage}

\begin{figure}
\centering
\includegraphics[angle=270,width=1\columnwidth]{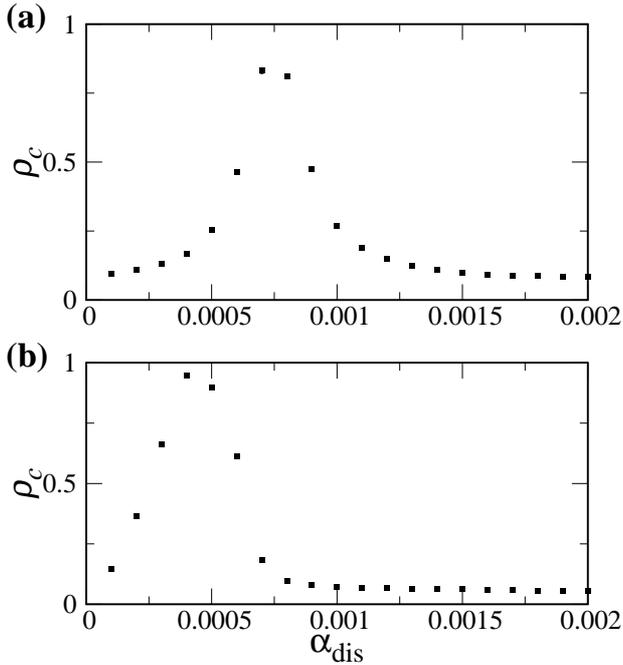}
\caption{Cooperation level $\rho_c$ as a function of the survival disadvantage parameter $\alpha_{\mathrm{dis}}$. In our present model, if the payoff $\pi_i$ of an agent $i$ disatisfies the minimum requirements ($\pi_i < M$), the extinction probability increases by $\alpha_{\mathrm{dis}}$. There is an optimal $\alpha_{\mathrm{dis}}$ for the promotion of high levels of cooperation; the value of the optimal cooperation-promoting $\alpha_{\mathrm{dis}}$ depends upon the approval time $\tau$: it is larger at shorter approval times, and smaller at longer approval periods. In both plots, we show the results for our generalized 6-strategy 2-tag model with time-varying tags, and the corresponding dominant strategy for the displayed results was always altrocentrism. (a) The approval time is $\tau = 26$. (b) The approval time is $\tau = 40$.}
\label{fig7}
\end{figure}

In our generalized 6-strategy model with time-varying tags, it is essential for agents surrounding a cooperative cluster to endure the extinction process until they have been approved to enter the cluster. Therefore, in addition to the approval time $\tau$, the fitness disadvantage parameter $\alpha_{\mathrm{dis}}$ is another crucial factor for the survival of agents. When $\alpha_{\mathrm{dis}}$ is low, agents who dissatisfy the minimum requirements may still survive for a rather long time. These agents disrupt the formation process of the cooperative cluster and the interconnectivity across cooperative clusters. As a result, a giant cooperative cluster cannot be formed. On the other hand, when $\alpha_{\mathrm{dis}}$ is too high, 
the agents surrounding the cooperative cluster cannot endure the extinction process; the cooperative cluster does not grow and ultimately 
vanishes. Consequently, for a given approval time, there is an optimal $\alpha_{\mathrm{dis}}$ which substantially enlarges cooperative clusters enabling the formation of a giant cluster. Fig.~\ref{fig7} confirms the existence of this optimal $\alpha_{\mathrm{dis}}$ necessary for attaining a high cooperation level 
at a given approval time. We see that the optimal value of $\alpha_{\mathrm{dis}}$ is inversely proportional to the approval time $\tau$. This is because 'new' agents surrounding cooperative clusters must endure more time if the associated approval period $\tau$ is long.  

\subsection{\label{sec:level8}General discussion and future research directions}

None of the previous cooperation models has ever studied the influence of heterogeneous naturalization approval times on cooperation under time-varying tag information. In the present study, we have found that our model with time-varying tags, heterogeneous naturalization duration, and multiple conditional strategies in an ageing population of artificial agents can give rise to highly rich dynamics that have not been reported previously. Perhaps most surprisingly, we have observed that under moderate approval times, cooperation in our model with time-varying tags subject to immigration dynamics can flourish even under very costly conditions that are otherwise highly detrimental to cooperative behavior. 

We have seen that cooperation in both the standard and in our generalized new model depended on $\tau$ and $c$, and that there was an interaction effect between these two variables on cooperation. In the standard 4-strategy model with fixed tags, cooperation was well-maintained only at lower values of $\tau$ and $c$. Above the critical cost $c_{\mathrm{crit}}$, cooperation in the standard model transitioned abruptly to a state of pure defection, and already at the lower range of intermediate approval times, cooperation was vanishingly attenuated without any signs of recovery at longer approval times. These results reveal remarkably novel insights into the fragility of ingroup-biased generosity, and corroborate the conclusions of more 
recent investigations~\cite{ramazi,hadzicuiwu,hadziliuli} highlighting the limits of ethnocentrism and its rather questionable robustness under a wide variety of realistic conditions.

On the other hand, the maintenance of altrocentric cooperation was observed in our new model throughout the range of approval times $\tau$ and values of the cost $c$, including the subregion of the parameter space above the critical cost $c_{\mathrm{crit}}$, where the level of altrocentric strategy first reached its peak and then gradually decreased, ultimately turning into the phase of dominant defection. Particularly high levels of cooperation in our 6-strategy model with time-varying tags were thus established at remarkably high costs $c_{\mathrm{crit}} < c < 0.1$, but only at intermediate $\tau$ (such as $20 < \tau < 39$) and not at low or high values of $\tau$. These findings thus suggest that instead of using some arbitrarily fixed (either short or long) approval times, there is an optimal duration of the naturalization procedure for new agents (that is of moderate length), from which the society as whole can profit most.

We note here that similar conditional strategies were previously employed in a model of discrimination emerging in spatial PD games with 
multiple tags~\cite{Jensen2019pre}; however, this model considered only a closed system of agents without immigration dynamics and without any approval mechanism. In addition, the strategies in this discrimination model were based on two fixed color traits that were hence not temporally variable.

In the context of cultural evolution, a recent model of trait change~\cite{pascualcuesta} investigated dynamic modifications of phenotypic features via imitation of other traits available in the population, whereby the underlying imitating behavior can be biased through homophily. Interestingly, the model allows for interaction among cultural traits which can further either reinforce or hinder each other via epistasis. Since this model has not been studied previously within the scope of evolutionary game theory, it would be challenging to investigate the effects of such epistatic cultural trait interactions on the evolution of strategies in our present tag-based cooperation model with time-varying phenotypic traits.  

In addition to the dynamics of time-varying phenotypic features, we suggest that subsequent extensions of our model should investigate the effects of immigration and naturalization procedures on cooperation in dynamic~\cite{WU2020124692} temporal networks~\cite{liding} with time-varying interactions~\cite{weisu}, 
where competing strategies, phenotypic features, and interaction structure can co-evolve at several different time-scales~\cite{rocacuestanxo}. Moreover, the behavior of such coevolutionary models should be investigated in the context of coupled information-cooperation processes on multilayer networks, where e.g. the evolutionary game dynamics is taking place in one network layer (as in our present paper), while in the other layer there is a spread of opinions~\cite{lichtenegger} about different immigration topics (including the debates on the duration of naturalization procedures); the opinion layer can then influence the game dynamics in the physical interaction layer, and conversely, the game outcomes can alter the evolution of opinions in the information layer, affecting thereby the population consensus about immigration and naturalization issues. In addition, other lattices~\cite{archimed} and other 
network types~\cite{whuycluster} should be explored as the underlying interaction structures of these multilayered systems. 

Beyond strategic multiplicity, when contacts are established among different groups of individuals such as native and non-native ones, different types of interactions or games are viable, such that one game is played preferentially within groups whereas another one is played with the outgroups~\cite{zgcaofashion}. We thus envision generalizations of our present work that would in future models include both multiple strategies and multiple games played within and across groups of players with time-varying tags that could dynamically change not only due to approval mechanisms but also via imitation dynamics among interacting agents. Here, the diversity of outgroups has to be carefully taken into consideration, as not all heterospecifics are necessarily viewed and evaluated in exactly the same way~\cite{faber,yangvanderdoes}. 

This research direction could further facilitate a more complete understanding of intergroup disharmony~\cite{lansing} and conflict~\cite{bravoyantseva}, which can significantly and sometimes even irreversibly impede cooperative behavior at a multitude of scales~\cite{dedeoflack}. In analogy to scale misperception phenomena in socio-economic systems~\cite{ostrometal}, it also needs to be acknowledged in future models that timescales of interactions among different populations as well as the underlying environmental cycles may not coincide with the timescale of decisions made about these populations, but may instead require strategic periodic behaviors that could play decisive roles in the management and prevention of collective conflict phenomena~\cite{dedeoperiodic}. 

\section{\label{sec:level8}Conclusions}
 
In this paper, we investigated the behavior of a novel evolutionary model of cooperation with time-varying phenotypic features, multiple conditional and unconditional strategies, and tag-mediated interactions occurring in a population with native and non-native agent dynamics subject to immigration and heterogeneous approval times. We asked if there exists an optimal period for the duration of naturalization procedures of newcomer agents, during which cooperation in the receiving society can reach its highest levels, even in the presence of high integration costs and a potentially emerging global defection. 

We found that in the standard 4-strategy model of tag-based cooperation with fixed traits, the ingroup-biased ethnocentric strategy can dominate only under short approval times and low costs of cooperation, revealing remarkable fragility of ethnocentric behavior. On the other hand, our generalized 6-strategy model with time-varying tags revealed that altrocentrism can outweigh all other competing strategies for a wider region of the parameter space, yielding the highest cooperation levels at intermediate approval times and at remarkably high costs that are usually detrimental to cooperative behavior. Additionally, we showed that the extinction probability of individual agents in our model is strongly related with the approval time. 

Our findings suggest that without relaxing the naturalization procedures or implementing any specially permissive immigration policies, high levels of social cooperation can be attained if the fraction of the population adopts the altrocentric strategy with an egalitarian generosity directed towards both native and approved naturalized citizens, regardless of their actual origin. These findings also suggest that instead of relying upon arbitrarily fixed approval times, there is an optimal duration of the naturalization procedure from which the society as a whole can profit most. Determining such optimal approval times in different social systems open to immigration may be paramount for the sustenance of large-scale cooperation. 

Importantly, the model developed in our present paper is relevant not only for understanding the effects of immigration on cooperation in socio-political systems such as nation states, but may also prove valuable for analyzing nearly any open system subject to invasion dynamics and increasing diversity such as economic markets (e.g. when new products or corporations enter the market, especially in the presence of conditional strategies such as consumer ethnocentrism), ecological systems (e.g. in the study of species invasiveness under habitat and resource scarcity), or artificial technological networks (e.g. in online peer-to-peer (P2P) systems, whenever new individuals are about to enter and use a file-sharing or a blockchain P2P network). We hope our present findings can stimulate further research in these directions as well as motivate innovative and more efficient approval policies in these systems, as they may become increasingly relevant in the face of a constantly changing and mobile world shattered by the ongoing pandemic crisis. 

\section*{Manuscript information}
The first version of this manuscript was completed on December 28, 2020, which was then revised on February 4 and again on March 15, 2021. The manuscript was then updated to its present form on March 29, 2021. 

\section*{Acknowledgments}
This work was supported by a GIST Research Institute (GRI) grant, funded by the Gwangju Institute of Science and Technology, South Korea, in 2020.

\bibliographystyle{elsarticle-num}

\onecolumn
\appendix
\section{The $12 \times 12$ payoff matrix for the PD game with six strategies and two tags}

The full $12 \times 12$ payoff matrix for our dynamic tag-based cooperation model with a six-strategy game structure and two tags can be written as $\mathbf{U} = [U_{pq}] =$

\footnotesize 
\begin{equation}
\hspace*{-0.6cm} 
\begin{array}{cc} 
&A_1~~~~~~~~E_1~~~~~~~~I_1~~~~~~~~~O_1~~~~~~~~N_1~~~~~~~~S_1~~~~~~~~A_2~~~~~~~~E_2~~~~~~~~I_2~~~~~~~~~O_2~~~~~~~~N_2~~~~~~~~S_2\\ \\
\begin{array}{c}
    A_1\\ \\
    E_1\\ \\
   I_1\\ \\
O_1\\ \\
N_1\\ \\
S_1\\ \\
A_2\\ \\
E_2\\ \\
I_2\\ \\
O_2\\ \\
N_2\\ \\
S_2\\
  \end{array}& \left(
  \begin{array}{cccccccccccc}
  u_{CC} &~ u_{CD} &~u_{CC} &~u_{CD}  &~ u_{CD}  &~u_{CC}  &~u_{CC}   &~ u_{CD} &~ u_{CD}  &~u_{CC}  &~u_{CD}   &~ u_{CC} \\ \\
    u_{DC} &~ u_{DD} &~u_{DC} &~u_{DD}  &~ u_{DD}  &~u_{DC}  &~u_{DC}   &~ u_{DD} &~ u_{DD}  &~u_{DC}  &~u_{DD}   &~ u_{DC} \\ \\
   u_{CC} &~ u_{CD} &~u_{CC} &~u_{CD}  &~ u_{CD}  &~u_{CC}  &~u_{DC}   &~ u_{DD} &~ u_{DD}  &~u_{DC}  &~u_{DD}   &~ u_{DC} \\ \\
   u_{DC} &~ u_{DD} &~u_{DC} &~u_{DD}  &~ u_{DD}  &~u_{DC}  &~u_{CC}   &~ u_{CD} &~ u_{CD}  &~u_{CC}  &~u_{CD}   &~ u_{CC} \\ \\
 u_{DC} &~ u_{DD} &~u_{DC} &~u_{DD}  &~ u_{DD}  &~u_{DC}  &~u_{CC}   &~ u_{CD} &~ u_{CD}  &~u_{CC}  &~u_{CD}   &~ u_{CC} \\ \\
 u_{CC} &~ u_{CD} &~u_{CC} &~u_{CD}  &~ u_{CD}  &~u_{CC}  &~u_{DC}   &~ u_{DD} &~ u_{DD}  &~u_{DC}  &~u_{DD}   &~ u_{DC} \\ \\
 u_{CC} &~ u_{CD} &~u_{CD} &~u_{CC}  &~ u_{CC}  &~u_{CD}  &~u_{CC}   &~ u_{CD} &~ u_{CC}  &~u_{CD}  &~u_{CC}   &~ u_{CD} \\ \\
 u_{DC} &~ u_{DD} &~u_{DD} &~u_{DC}  &~ u_{DC}  &~u_{DD}  &~u_{DC}   &~ u_{DD} &~ u_{DC}  &~u_{DD}  &~u_{DC}   &~ u_{DD} \\ \\
 u_{DC} &~ u_{DD} &~u_{DD} &~u_{DC}  &~ u_{DC}  &~u_{DD}  &~u_{CC}   &~ u_{CD} &~ u_{CC}  &~u_{CD}  &~u_{CC}   &~ u_{CD} \\ \\
 u_{CC} &~ u_{CD} &~u_{CD} &~u_{CC}  &~ u_{CC}  &~u_{CD}  &~u_{DC}   &~ u_{DD} &~ u_{DC}  &~u_{DD}  &~u_{DC}   &~ u_{DD} \\ \\
 u_{DC} &~ u_{DD} &~u_{DD} &~u_{DC}  &~ u_{DC}  &~u_{DD}  &~u_{CC}   &~ u_{CD} &~ u_{CC}  &~u_{CD}  &~u_{CC}   &~ u_{CD} \\ \\
 u_{CC} &~ u_{CD} &~u_{CD} &~u_{CC}  &~ u_{CC}  &~u_{CD}  &~u_{DC}   &~ u_{DD} &~ u_{DC}  &~u_{DD}  &~u_{DC}   &~ u_{DD} \\
\end{array}
\right)\\
\end{array} 
\end{equation}

\normalsize
\medskip
\bigskip

\noindent whereby $U_{pq}$ is the resulting payoff of an individual player with a given strategy ($A$, $E$, $I$, $O$, $N$, or $S$) and a tag denoted with a subscript number $1$ ('approved') or $2$ ('new') of row $p$ when interacting with an opponent player employing the strategy and the tag of 
column $q$ ($A$=altruism, $E$=egoism, $I$=ethnocentrism, $O$=cosmopolitanism, $N$=neophilia, and $S$=sympolitic altrocentrism). Considering the two-player game design employed in our present model, $u_{CC}=b - c$ denotes the payoff that is distributed to players for their joint cooperation (C), such that both co-players receive the benefit $b$ and incur the cost $c$ for their mutual generosity. On the other hand, mutual defection (D) results in the payoff $u_{DD}=0$. If one of the two interacting players decides to donate help to the opponent while the other one refuses to cooperate and defects, the cooperator incurs a cost $c$ without obtaining any benefit whatsoever, resulting thereby in a payoff $u_{CD}=- c$; the defector, in contrast, benefits from this interaction without incurring any costs with a payoff $u_{DC}=b$ (whereby the payoff ranking is given by $u_{DC}>u_{CC}>u_{DD}>u_{CD}$). The detailed payoff matrix for our 6-strategy game structure with two phenotypic features (tags) can therefore be rewritten as $\mathbf{U} =$
\smallskip

\small
\begin{equation}
\hspace*{-0.56cm} 
\begin{array}{cc} 
&A_1~~~~~~~~~~E_1~~~~~~~~~~I_1~~~~~~~~~O_1~~~~~~~~N_1~~~~~~~~~S_1~~~~~~~~A_2~~~~~~~~~~E_2~~~~~~~~~~I_2~~~~~~~~~O_2~~~~~~~~~~N_2~~~~~~~~~S_2\\ \\
\begin{array}{c}
    A_1\\ \\
    E_1\\ \\
   I_1\\ \\
O_1\\ \\
N_1\\ \\
S_1\\ \\
A_2\\ \\
E_2\\ \\
I_2\\ \\
O_2\\ \\
N_2\\ \\
S_2\\
  \end{array}& \left(
  \begin{array}{cccccccccccc}
  b-c &~ -c &~b-c &~-c  &~ -c  &~b-c  &~b-c   &~ -c  &~ -c  &~b-c  &~-c   &~ b-c\\ \\
    b &~ 0 &~b &~0  &~ 0  &~b  &~b   &~ 0  &~ 0  &~b  &~0   &~ b \\ \\
   b-c &~ -c &~b-c &~-c  &~ -c  &~b-c  &~b   &~ 0  &~ 0  &~b  &~0   &~ b \\ \\
   b &~ 0 &~b &~0  &~ 0  &~b  &~b-c   &~ -c &~ -c  &~b-c  &~-c   &~ b-c \\ \\
 b &~ 0 &~b &~0  &~ 0  &~b  &~b-c   &~ -c  &~ -c  &~b-c  &~-c   &~ b-c \\ \\
 b-c &~ -c &~b-c &~-c  &~ -c  &~b-c  &~b   &~ 0  &~ 0  &~b  &~0   &~ b \\ \\
 b-c &~ -c &~-c &~b-c  &~ b-c  &~-c  &~b-c   &~ -c  &~ b-c  &~-c  &~b-c   &~ -c\\ \\
  b &~ 0 &~0 &~b  &~ b  &~0  &~b   &~ 0 &~ b  &~0  &~b   &~ 0 \\ \\
 b &~ 0 &~0 &~b  &~ b  &~0  &~b-c   &~ -c  &~ b-c  &~-c  &~b-c   &~ -c \\ \\
 b-c &~ -c &~-c &~b-c  &~ b-c  &~-c  &~b   &~ 0  &~ b  &~0  &~b   &~ 0 \\ \\
 b &~ 0 &~0 &~b  &~ b  &~0  &~b-c   &~ -c  &~ b-c  &~-c  &~b-c   &~ -c\\ \\
 b-c &~ -c &~-c &~b-c  &~ b-c  &~-c  &~b   &~ 0  &~ b  &~0  &~b   &~ 0 \\
\end{array}
\right)\\
\end{array} 
\end{equation}

\end{document}